\documentclass{tMPH2e}
\usepackage{amssymb}
\usepackage{makeidx}
\usepackage{amsmath}
\usepackage{graphicx}
\usepackage{epsfig}
\usepackage[usenames]{color}
\usepackage{subfigure}

\newcommand{\beq}{\begin{equation}}
\newcommand{\eeq}{\end{equation}}
\newcommand{\beqa}{\begin{eqnarray}}
\newcommand{\eeqa}{\end{eqnarray}}
\newcommand{\ba}{\begin{array}}
\newcommand{\ea}{\end{array}}

\begin{document}

\title{Spontaneous symmetry breaking in linearly coupled \\
disk-shaped Bose-Einstein condensates}
\author{Luca Salasnich$^{1}$ and Boris A. Malomed$^{2,3}$ \\
$^{1}$Dipartimento di Fisica ``Galileo Galilei'' and CNISM, Universit\`a di
Padova, \\
Via Marzolo 8, 35131 Padova, Italy \\
$^{2}$Department of Physical Electronics, School of Electrical Engineering,
Faculty of Engineering, Tel Aviv University, Tel Aviv 69978, Israel\\
$^3$ICFO-Institut de Ciencies Fotoniques, Mediterranean Technology Park,
08860 Castelldefels (Barcelona), Spain$^*$\thanks{$^*$temporary Sabbatical
address}}
\maketitle

\begin{abstract}
We study effects of tunnel coupling on a pair of parallel disk-shaped
Bose-Einstein condensates with the self-attractive intrinsic nonlinearity.
Each condensate is trapped in a combination of in-plane and
transverse harmonic-oscillator potentials. It is shown that, depending on
the self-interaction strength and tunneling coupling, the ground state of
the system exhibits a phase transition which links three configurations: a
symmetric one with equal numbers of atoms in the coupled condensates, an
asymmetric configuration with a population imbalance (a manifestation of the
macroscopic quantum self-trapping), and the collapsing state. A modification
of the phase diagram of the system in the presence of vortices in the
disk-shaped condensates is reported too. The study of dynamics around the
stationary configurations reveals properties which strongly depend on the
symmetry of the configuration.
\end{abstract}

\doi{10.1080/0026897YYxxxxxxxx} \issn{13623028} \issnp{00268976} %
\jvol{00} \jnum{00} \jyear{2011} 

\markboth{Taylor \& Francis and I.T. Consultant}{Molecular Physics}

\section{Introduction}

It has been predicted in many theoretical works \cite{bifurcation} that two
dilute symmetric Bose-Einstein condensates (BECs), which are weakly coupled
by the tunneling of atoms across the separating potential barrier, can give
rise to the macroscopic quantum self-trapping. In particular, in the case of
attractive inter-atomic interactions, the ground state of the system shows a
transition from the symmetric configuration (the Josephson regime),
characterized by equal numbers of atoms in the coupled condensates, to an
asymmetric state (the self-trapping regime), characterized by an imbalance
in the number of atoms \cite{bifurcation,leggett}. In the case of repulsive
interactions, the self-trapping occurs not in the ground state , but rather
in the first antisymmetric excited state. In the latter case, the
self-trapping was demonstrated in experiments with the condensate of $^{87}$%
Rb atoms \cite{Heidelberg} (for a review, see Ref. \cite{Oliver}).

The self-trapping in the BEC loaded into the double-well potential is a
manifestation of the general effect of the spontaneously symmetry breaking
(SSB) in nonlinear systems. As said above, asymmetric states trapped in
symmetric potentials are generated by SSB \textit{bifurcations} from obvious
symmetric or antisymmetric states, in the media with the attractive or
repulsive intrinsic nonlinearity, respectively (the SSB under the action of
competing attractive (cubic) and repulsive (quintic) terms was studied too,
featuring closed \textit{bifurcation loops} \cite{CQ,Nir}). In terms of BEC
and other macroscopic quantum systems, the SSB may also be realized as a
phase transition, which replaces the original symmetric ground state  by a
new asymmetric one, when the strength of the self-attractive nonlinearity
exceeds a certain critical value. A transition of this type was actually
predicted earlier in classical systems, \textit{viz}., in a model of
dual-core nonlinear optical fibers with the self-focusing Kerr nonlinearity
\cite{dual-core}. In connection to the interpretation of the SSB as the
phase transition, it may be identified as the transition of the first or
second kind (alias \textit{subcritical} or \textit{supercritical} type of
the SSB bifurcation), depending on the form of the nonlinearity, spatial
dimension, and the presence or absence of an external periodic potential (an
optical lattice) acting along the additional spatial dimension (if any) \cite%
{Warsaw,Arik2}.

Theoretical studies of the SSB in BECs were extended in various directions,
especially for matter-wave solitons. In particular, the symmetry breaking of
the solitons was predicted in various two-dimensional (2D) settings \cite%
{Warsaw}, including the spontaneous breaking of the skew symmetry of
solitons and localized vortices trapped in double-layer condensates with
mutually orthogonal orientations of quasi-one-dimensional optical lattices
induced in the two layers \cite{skew}. A different variety of the 2D
geometry, which gives rise to its own mode of the SSB, is based on a
symmetric set of four potential wells \cite{4wells}. Self-trapping of
asymmetric states was also predicted in condensates formed of dipolar atoms,
which interact via long-range forces \cite{DD}, and in the context of the
nonlinear Schr\"{o}dinger equation with a general local nonlinearity \cite%
{higher-order}. The symmetry breaking is possible not only in linear
potentials composed of two wells, but also in a similarly structured \textit{%
pseudopotentials}, which are produced by a symmetric spatial modulation of
the non-linearity coefficient, with two sharp maxima \cite{Dong,Barcelona}.

Another generalization was developed for the SSB in two- \cite{2comp} and
three-component (spinor) \cite{3comp} BEC mixtures, where the asymmetry of
the density profiles in the two wells comes along with a difference in
distributions of the different species. As concerns multi-component systems,
the analysis of the SSB was also extended to Bose-Fermi mixtures \cite%
{Kivshar}.

On the other hand, it is commonly known that the self-attraction in BEC may
cause collapse of the condensate in the form of a \textquotedblleft
bosenova" (in which case three-body recombinations become important, in
addition to the usual two-body collisions \cite{bosenova}). Therefore, the
SSB in BEC trapped in double-well (dual-core) potential may compete with the
collapse. Recently, we have determined \cite{we1,we2} the domain of
parameters of such a symmetric dual-core system above which the collapse
occurs. In particular, the competition between the SSB and the onset of the
collapse in a pair of parallel cigar-shaped atomic condensates weakly
coupled by tunneling of atoms was investigated in Ref. \cite{we1}. Further,
in Ref. \cite{we2}, the SSB and collapse were studied in a quasi-1D bosonic
Josephson junction made by a double-well potential in the axial direction,
and by a harmonic potential in the radial directions.

In the present paper we consider a different setup, namely, a pair of
parallel disk-shaped atomic condensates weakly-coupled by tunneling of atoms
and confined by harmonic-oscillator potentials. This setup is ideal
to analyze the interplay of the nonlinearity and tunnel coupling in the
presence of vortices in both condensates \cite{Arik2}. In contrast to Ref.
\cite{Arik2}, which described this system by a pair of linearly-coupled 2D
Gross-Pitaevskii equations (GPEs) with the cubic nonlinearity, and actually
presented the analysis of the SSB only below the collapse threshold, in this
work we use the more accurate system of equations with the nonpolynomial
nonlinearity, specific to the 2D geometry \cite{nonpoly}, and the
competition of the SSB with the onset of the collapse is one of main goals.
After formulating the model in Section 2, we consider, in Section 3, the
ground state of the system, which shows a phase transition between three
possible configurations: a symmetric one, with equal numbers of atoms in the
two coupled condensates, an asymmetric configuration with a population
imbalance (the macroscopic self-trapping, in the present setting), and the
collapsing state. Then, we perform a similar analysis for localized states
carrying vorticity in each core, which changes the phase diagram of the
system. Finally, we study the dynamics of the two disk-shaped condensates
around the stationary configurations. Starting from a symmetric
configuration, we predict small-amplitude Josephson-like oscillations, with
periodic transfer of the population imbalance from one core to the other.
Starting from an asymmetric configuration, we find, instead, large-amplitude
oscillations, which preserve the population imbalance. The paper is
concluded by summary and discussion of open problems in Section 4.

\section{The model}

\subsection{The dimensional reduction from 3D to 2D}

The starting point is the three-dimensional GPE for the mean-field wave
function, $\psi (\mathbf{r},t)$, which describes BEC in two parallel
identical disk-shaped traps separated by a potential barrier:
\begin{eqnarray}
i\hbar {\frac{\partial }{\partial t}}\psi &=&\frac{1}{2}\left\{ -\frac{\hbar
}{m}\nabla ^{2}+m\omega _{z}^{2}[(z-z_{0})^{2}+(z+z_{0})^{2}]\right\} \psi
\notag \\
&+&W(x,y)\psi +\frac{4\pi \hbar ^{2}a_{s}}{m}\left\vert \psi \right\vert
^{2}\psi ,  \label{GPEs}
\end{eqnarray}%
where $z$ is the coordinate transversal to the disks, $2z_{0}$ is the
separation between their centers along the $z$-direction, the two harmonic
potentials with frequency $\omega _{z}$ account for the transverse trapping
of atoms in each disk, and $W(x,y)$ is the potential acting in the disk
plane (it is assumed to be identical for both disks). As usual, $a_{s}$ is
the $s$-wave inter-atomic scattering length \cite{GP}.

The first objective is to reduce Eq. (\ref{GPEs}) to a system of linearly
coupled equations for 2D wave functions pertaining to the separate disks, $%
\Phi _{1,2}$. To this end, we modify the approach developed for the system
of two parallel quasi-1D \textquotedblleft cigars" in Ref. \cite{we1},
adopting a superposition of two single-disk \textit{ans\"{a}tze}:%
\begin{eqnarray}
\psi (\mathbf{r},t) &=&\pi ^{-1/4}\left[ \exp \left( -\frac{\left(
z-z_{0}\right) ^{2}}{2\eta _{1}(x,y,t)^{2}}\right) \frac{\Phi _{1}(x,y,t)}{%
\sqrt{\eta _{1}(x,y,t)}}\right.   \notag \\
&&\left. +\exp \left( -\frac{\left( z+z_{0}\right) ^{2}}{2\eta
_{2}(x,y,t)^{2}}\right) \frac{\Phi _{2}(x,y,t)}{\sqrt{\eta _{2}(x,y,t)}}%
\right] ,  \label{ansatz}
\end{eqnarray}%
where $\eta _{1}(x,y,t)$ and $\eta _{2}(x,y,t)$ are the thicknesses of the
two disks along the $z$ axis, and the 1D part of each wave function is
normalized to unity.

We proceed by substituting ansatz (\ref{ansatz}) into the Lagrangian
corresponding to Eq. (\ref{GPEs}),
\begin{gather}
L=\frac{1}{2}\int d^{3}\mathbf{r}\Big\{i\hbar \Big({\frac{\partial \psi
^{\ast }}{\partial t}}\psi -{\frac{\partial \psi }{\partial t}}\psi ^{\ast }%
\Big)+\frac{\hbar ^{2}}{m}|\nabla \psi |^{2}  \notag \\
+m\omega _{z}^{2}[(z-z_{0})^{2}+(z+z_{0})^{2}]|\psi |^{2}  \notag \\
+2W(x,y)|\psi |^{2}+\frac{4\pi \hbar ^{2}a_{s}}{m}|\psi |^{4}\Big\}:.
\label{L}
\end{gather}%
The underlying assumption is that distance $2z_{0}$ between the disks is
essentially larger than the size $a_{z}=\sqrt{\hbar /(m\omega _{z})}$ of the
transverse confinement in each of them, $2z_{0}\gg a_{z}$. Due to this
condition, the part of the Lagrangian, which accounts for the tunneling and
is produced by the overlap of the two components of the wave function in
ansatz (\ref{ansatz}), if substituted into Lagrangian (\ref{L}), takes the
following form:
\begin{equation*}
L_{\mathrm{T}}=-K\int \left[ \Phi _{1}(x,y)\Phi _{2}^{\ast }(x,y)+\Phi
_{1}^{\ast }(x,y)\Phi _{2}(x,y)\right] dxdy,
\end{equation*}%
where the effective coupling coefficient is defined as
\begin{equation}
K=\hbar \omega _{z}{\frac{z_{0}^{2}}{a_{z}^{2}}}\exp \left( -\frac{z_{0}^{2}%
}{a_{z}^{2}}\right) .  \label{kappa2}
\end{equation}%
In fact, the main contribution to the linear coupling (tunneling) comes from
region $z^{2}\lesssim \eta _{1,2}^{2}$ around the \emph{midpoint} between
the disks. In that region, the transverse-confinement radius is determined
by the ground-state wave function of the 1D harmonic oscillator, which has
characteristic length $a_{z}$ in the $z$ direction.

Finally, the effective dynamical equations for the two linearly coupled
disks ($n=1,2)$ are written as
\begin{eqnarray}
i{\frac{\partial }{\partial t}}\Phi _{n} &=&\Big[-{\frac{1}{2}}\nabla _{\bot
}^{2}+W(x,y)+g{\frac{|\Phi _{n}|^{2}}{\eta _{n}}}  \notag \\
&&+{\frac{1}{4}}\left( {\frac{1}{\eta _{n}^{2}}}+\eta _{n}^{2}\right) \Big]%
\Phi _{n}-\kappa \ \Phi _{3-n}\;,  \label{npse}
\end{eqnarray}%
where the scaled interaction strength is $g\equiv \sqrt{2\pi }\gamma $, with
\begin{equation}
\gamma =2Na_{s}/a_{z},  \label{gamma}
\end{equation}%
the scaled linear coupling is
\begin{equation}
\kappa =K/(\hbar \omega _{z}),  \label{kappa}
\end{equation}%
and the respective axial widths are determined by algebraic equations,
\begin{equation}
\eta _{n}^{4}=1+g|\Phi _{n}|^{2}\eta _{n},\;.  \label{eta}
\end{equation}%
Notice that in Eqs. (\ref{npse}) and (\ref{eta}) we have used scaled
variables, \textit{viz}., the length measured in units of $a_{z}=\sqrt{\hbar
/(m\omega _{z})}$, time in units of $\omega _{z}^{-1}$, and energy in units
of $\hbar \omega _{z}$.

Exact solutions to Eqs. (\ref{eta}) can be found by way of the Cardano
formula,
\begin{eqnarray}
\eta _{n} &=&\pm {\frac{1}{2}}\sqrt{\frac{A_{n}^{2}-12}{3A_{n}}}  \notag \\
&+&{\frac{1}{2}}\sqrt{-{\frac{A_{n}^{2}-12}{3A_{n}}}\pm 2g|\Phi
_{n}|^{2}\left( \frac{A_{n}^{2}-12}{3A_{n}}\right) ^{-1/2}}\;,
\label{eta-solve}
\end{eqnarray}%
where the upper and lower signs correspond, respectively, to $g>0$ and $g<0$%
, and
\begin{equation}
A_{n}\equiv \left( {3/2}\right) ^{1/3}\left( 9g^{2}|\Phi _{n}|^{4}+\sqrt{3}%
\sqrt{256+27g^{4}|\Phi _{n}|^{8}}\right) ^{1/3}\;.  \label{A}
\end{equation}

\subsection{Properties of the 2D model}

Thus, we have reduced the initial 3D problem, based on Eq. (\ref{GPEs}), to
the 2D problem for the set of wave functions of the BECs trapped in the two
disks, which obey Eqs. (\ref{npse}) and (\ref{eta}). To stabilize 2D
solitons and vortices, we choose the in-plane potential as that of the 2D
harmonic oscillator,
\begin{equation}
W\left( x,y\right) =\frac{1}{2}\lambda ^{2}(x^{2}+y^{2}),  \label{W}
\end{equation}%
where $\lambda $ is the adimensional frequency of the planar confinement.
The system conserves the total number of atoms in the two disks, i.e., $%
N_{1}(t)+N_{2}(t)=2\ $(in the scaled units), where
\begin{equation*}
N_{n}=\int |\Phi _{n}(x,y,t)|^{2}dxdy\quad \quad (n=1,2)\;.
\end{equation*}%
Also conserved are the total energy and angular momentum.

It is relevant to mention that the (effectively) two-dimensional
self-attractive condensate, trapped in a periodic optical-lattice potential,
features not only the collapse, when its norm exceeds the corresponding
critical value, but also delocalization, when the norm falls below a certain
threshold (the latter effects is also known in other 
dimensions) \cite{Mario}. 
In the present setting, the delocalization does not occur, as we consider
the situation with the harmonic-oscillator potentials confining the
condensates in all directions.

Vortex-soliton solutions to Eqs. (\ref{npse}) are sought for as
\begin{equation*}
\Phi _{n}(r,\theta ,t)=\phi _{n}(r,t)\ e^{iS\theta }\;,
\end{equation*}%
where $r$ and $\theta $ are the polar coordinates in the $\left( x,y\right) $
plane, and $S$ is the integer vorticity. In this way, Eqs. (\ref{npse}) can
be reduced to coupled nonpolynomial Schr\"{o}dinger equations in the radial
direction,
\begin{eqnarray}
i{\frac{\partial }{\partial t}}\phi _{n} &=&\frac{1}{2}\Big[-\left( {\frac{%
\partial ^{2}}{\partial r^{2}}}+{\frac{1}{r}}{\frac{\partial }{\partial r}}%
\right) +{\frac{S^{2}}{r^{2}}}+\lambda ^{2}r^{2}+2g{\frac{|\phi _{n}|^{2}}{%
\eta _{n}}}  \notag \\
&+&{\frac{1}{2}}\left( {\frac{1}{\eta _{n}^{2}}}+\eta _{n}^{2}\right) \Big]%
\phi _{n}-\kappa \ \phi _{3-n}~.  \label{npse-cyl}
\end{eqnarray}%
Further, stationary states are then obtained by setting $\phi
_{n}(r,t)=u_{n}(r)\ e^{-i\mu _{n}t},$ with real functions $u_{n}$, which
obey a system of stationary radial equations,
\begin{eqnarray*}
\mu _{n}u_{n} &=&\frac{1}{2}\Big[-\left( {\frac{\partial ^{2}}{\partial r^{2}%
}}+{\frac{1}{r}}{\frac{\partial }{\partial r}}\right) +{\frac{S^{2}}{r^{2}}}%
+\lambda ^{2}r^{2}+2g{\frac{u_{n}^{2}}{\eta _{n}}} \\
&+&{\frac{1}{2}}\left( {\frac{1}{\eta _{n}^{2}}}+\eta _{n}^{2}\right) \Big]%
u_{n}-\kappa u_{3-n}\;,
\end{eqnarray*}%
while widths $\eta _{1,2}$ are still determined by Eqs. (\ref{eta}), with $%
\left\vert \Phi _{1,2}\right\vert ^{2}$ replaced by $u_{1,2}^{2}$.

As said above, the main objective of the work is to predict the SSB of the
symmetric solitons, with $u_{1}\left( r\right) =u_{2}\left( r\right) $, $%
\eta _{1}=\eta _{2}$. In the system of two linearly coupled GPEs with the
usual cubic nonlinearity, this problem was studied in Ref. \cite{Arik2}.
Here, we seek for stationary solutions by means of direct simulations of the
time-dependent cylindrically-symmetric coupled 2D equations (\ref{npse-cyl}%
), using a finite-difference Crank-Nicholson algorithm in the imaginary time
\cite{sala-numerics}. The initial conditions were taken as
\begin{equation}
\Phi _{n}(r,t=0)=C_{n}\ r^{S}\ \exp \left( -\lambda r^{2}/2\right) ,
\label{pippo}
\end{equation}%
where $C_{n}$ ($n=1,2$) are normalization constants. Note that Eq. (\ref%
{pippo}) gives the exact quantum-mechanical wave function of the stationary
vortex configuration in the absence of the nonlinearity ($g=0$) and linear
coupling ($\kappa =0$). In our numerical simulations we choose $C_{1}\neq
C_{2}$, but with $C_{1}$ taken very close to $C_{2}$, to initiate the
development of the symmetry breaking, if it possible. In particular, the
norms of functions $\Phi _{1}(r,t=0)$ and $\Phi _{2}(r,t=0)$ are taken as $%
1.01$ and $0.99$, respectively.

\section{Numerical results}

\subsection{The ground state and vortices}

\begin{figure}[tbp]
\center\includegraphics [width=10.cm,clip]{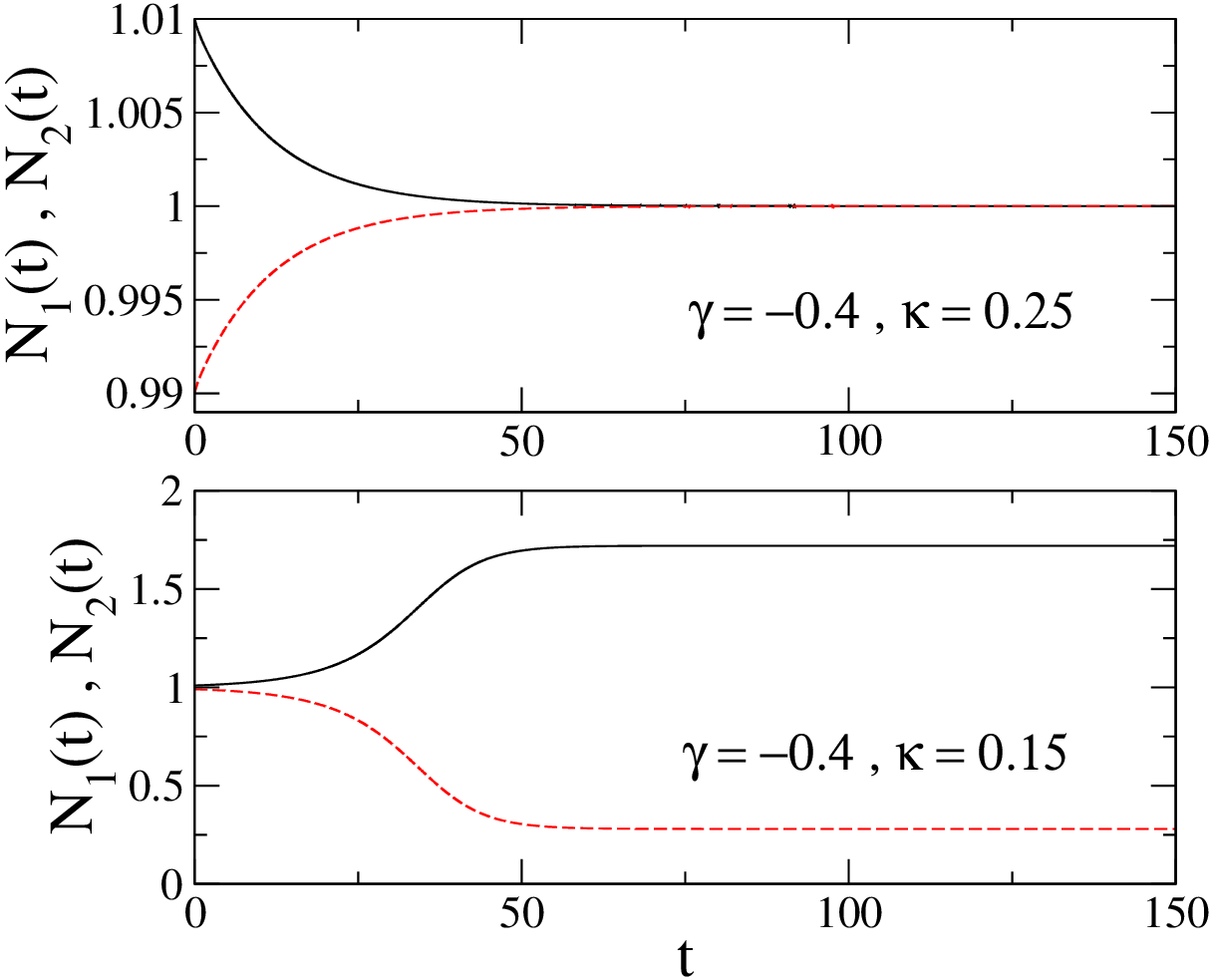}
\caption{(Color online) Norms $N_{1}$ and $N_{2}$ (solid and dashed lines)
of the ground (zero-vorticity, $S=0$) state trapped in the two parallel
disks, in the course of the evolution in imaginary time. The nonlinearity
coefficient is $\protect\gamma =-0.4$, while the coupling constant is $%
\protect\kappa =0.25$ (upper panel) and $\protect\kappa =0.15$ (lower
panel). The adimensional notation corresponds to the length measured in
units of $a_{z}=\protect\sqrt{\hbar /(m\protect\omega _{z})}$, time in units
of $\protect\omega _{z}$, and energy in units of $\hbar \protect\omega _{z}$%
. Recall that $\protect\gamma \ $and $\protect\kappa \ $are defined by Eqs. (%
\protect\ref{gamma}) and (\protect\ref{kappa}), respectively.}
\label{fig1}
\end{figure}

In Fig. \ref{fig1} we plot the norms $N_{1}$ and $N_{2}$ (solid and dashed
lines) of the two coupled condensates, with zero vorticity, $S=0$, in the
course of the evolution of in the imaginary time, by choosing $\gamma =-0.4$
[recall $\gamma $ is defined in Eq. (\ref{gamma}), $\gamma <0$ corresponding
to the attractive interatomic interactions], and slightly asymmetric initial
conditions (i.e. slightly imbalanced populations). As shown in the figure,
with $\kappa =0.25$ (the upper panel) the symmetry is restored during the
time evolution, while with $\kappa =0.15$ (the lower panel) the asymmetry is
strongly enhanced towards a finite population imbalance.

\begin{figure}[tbp]
\center\includegraphics [width=11.cm,clip]{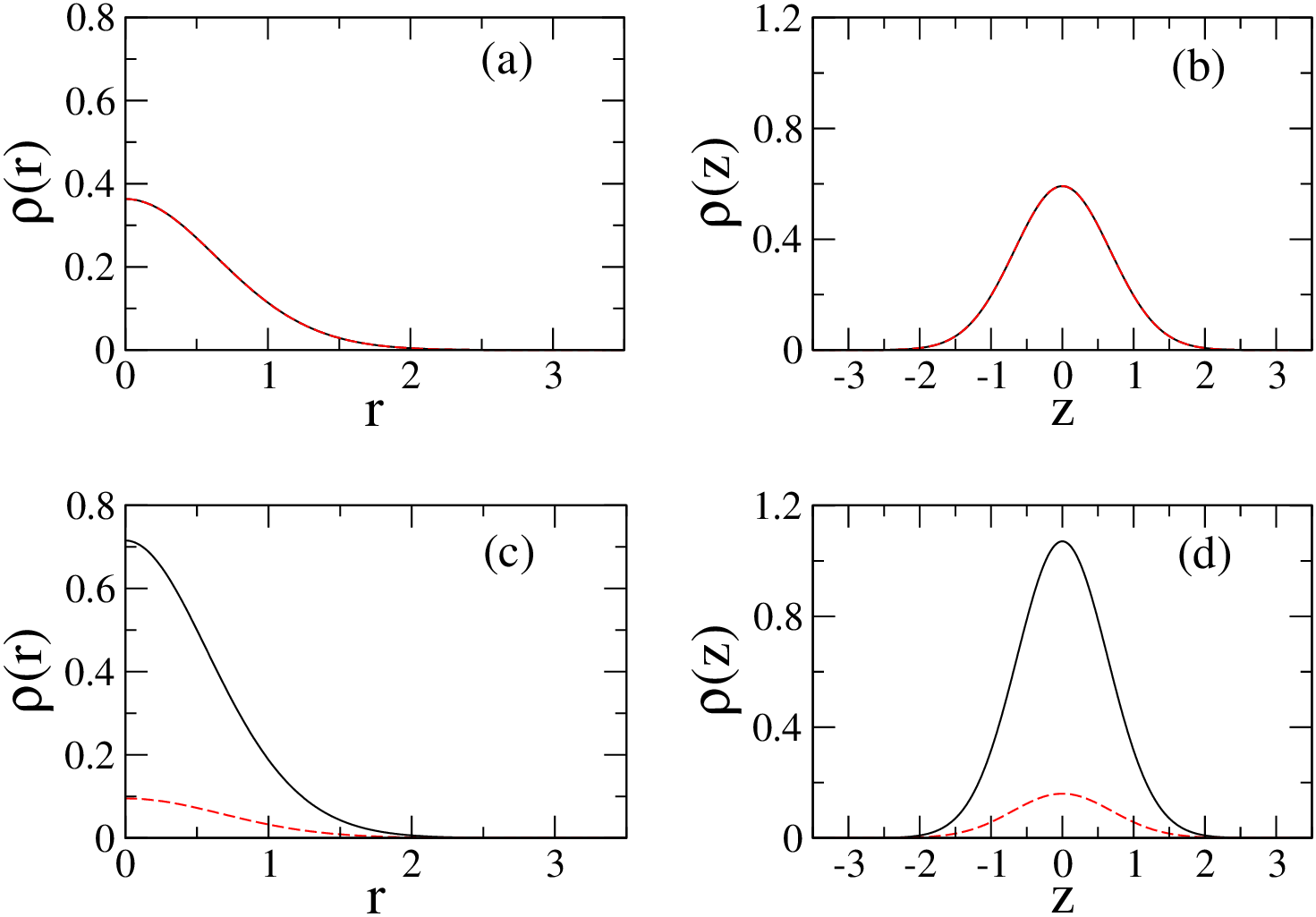}
\caption{(Color online) Stationary density profiles of the ground state ($S=0
$) in the two disks (solid and dashed lines), corresponding to the cases
shown in Fig. (\protect\ref{fig1}). The upper panels: radial density
profiles $\protect\rho (r)$ (a), and axial density profiles $\protect\rho (z)
$, see Eq. (\protect\ref{rho}), (b), for $\protect\gamma =-0.4$ and $\protect%
\kappa =0.25$. Lower panels: the radial profiles (c) and the axial profiles
(d) for $\protect\gamma =-0.4$ and $\protect\kappa =0.15$. Units are the
same as in Fig. \protect\ref{fig1}.}
\label{fig2}
\end{figure}

In the framework of the 2D description, the factorized ansatz (\ref{ansatz})
yields the time-dependent radial density profile, $\rho _{n}(r,t)=|\Phi
_{n}(r,t)|^{2},$ and its axial counterpart,
\begin{equation}
\rho _{n}(z,t)=2\sqrt{\pi }\int_{0}^{\infty }\frac{rdr}{\eta _{n}(r,t)}%
e^{-z^{2}/\eta _{n}(r,t)^{2}}|\Phi _{n}(r,t)|^{2}\;.  \label{rho}
\end{equation}%
In Fig. \ref{fig2} the corresponding final (stationary) density profiles are
displayed in the two disks, by solid and dashed lines. In the upper panels,
the density profiles of the symmetric state are fully superimposed, while in
the lower panels they are clearly distinguishable, for the asymmetric mode.

\begin{figure}[th]
\centering
\epsfig{file=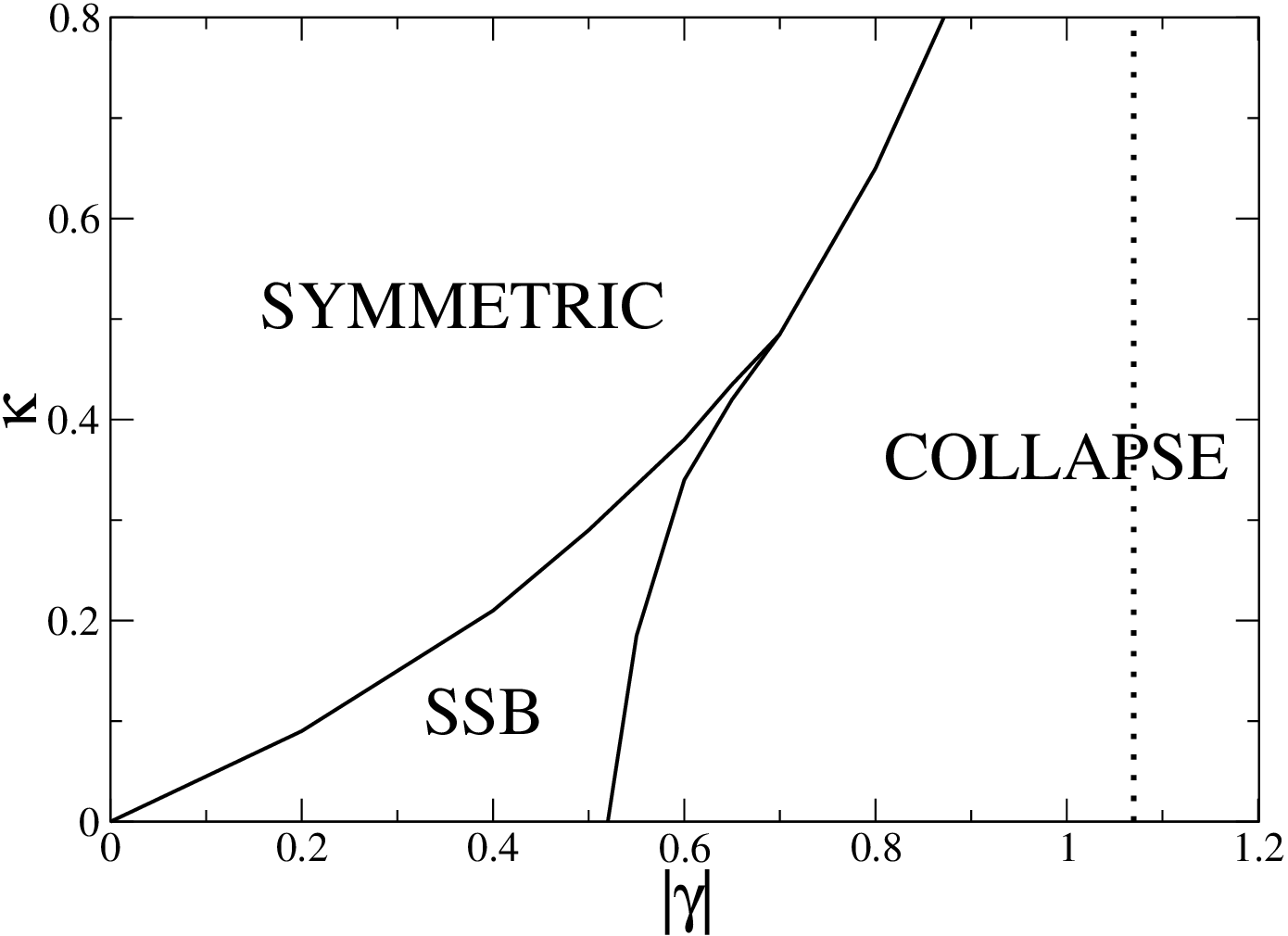,width=0.8\linewidth,clip=}
\caption{The phase diagram of the linearly-coupled system of 2D equations (%
\protect\ref{npse-cyl}) with $S=0$ (the ground state). In regions
\textquotedblleft symmetric" and \textquotedblleft SSB", the system
supports, respectively, stable symmetric [$u_{1}(r)=u_{2}(r)$] and
asymmetric [$u_{1}(r)\neq u_{2}(r)$] stationary solutions. The collapse
takes place in the eponymous region. The system always suffers the collapse
to the right of the vertical dashed line. Units are as in Fig. \protect\ref%
{fig1}.}
\label{fig3}
\end{figure}

Results of a systematic analysis, generated by varying parameters $\gamma $
and $\kappa $, are summarized in Fig. \ref{fig3}. Here we show the phase
diagram generated by the linearly-coupled system of 2D equations (\ref%
{npse-cyl}), with $S=0$, in the parameter plane. As explained also in the
the caption to the figure, in regions \textquotedblleft symmetric" and
\textquotedblleft SSB" the system supports, respectively, stable symmetric
and asymmetric stationary solutions. In region \textquotedblleft collapse",
the imaginary-time dynamics evolves towards a configuration with a
zero-length axial width (in one or both disks). Notice that, on the right
side of the vertical dashed line in Fig. \ref{fig3} (i.e., at $|\gamma |>1.07
$), the system always suffers the collapse, as in that region the
nonlinearity strength exceeds the critical value leading to the onset of the
collapse.

\begin{figure}[tbp]
\center\includegraphics [width=11.cm,clip]{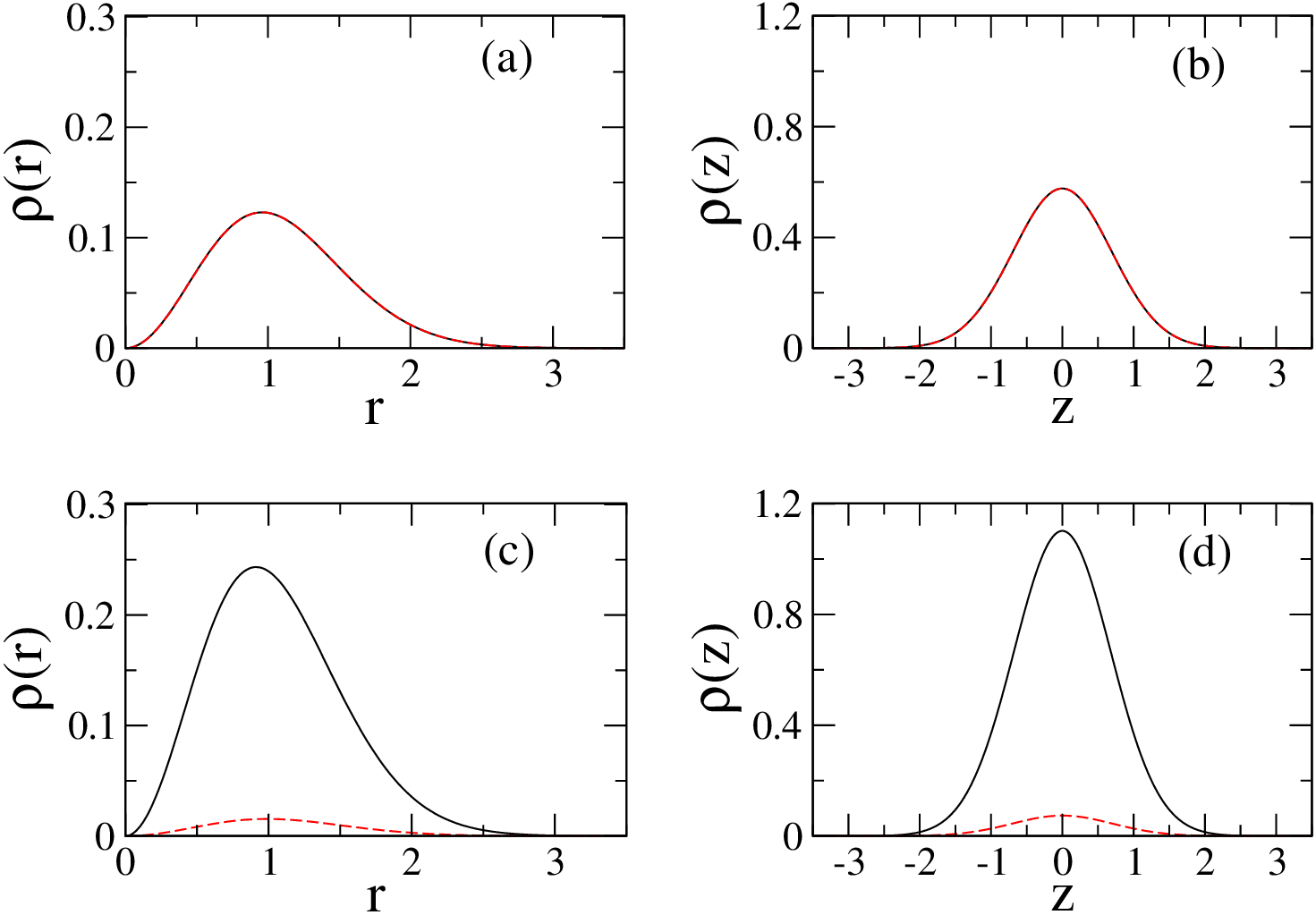}
\caption{(Color online) Stationary density profiles of vortices with $S=1$
in the two disks (solid and dashed lines). Upper panels: (a) radial density
profiles $\protect\rho (r)$ (a) and axial density profiles $\protect\rho (z)$
(b) for $\protect\gamma =-0.4$ and $\protect\kappa =0.2$. Lower panels:
radial profiles (c) and axial profiles (d) for $\protect\gamma =-0.05$ and $%
\protect\kappa =0.15$. The units are as in Fig. \protect\ref{fig1}.}
\label{fig4}
\end{figure}

The density profiles of the vortical states with $S=1$ in both disks are
displayed in Fig. \ref{fig4}, by means of the solid and dashed lines. The
left panels of the figure clearly show the impact of the vorticity on the
radial profiles $\rho (r)$, which vanish at $r\rightarrow 0$. The figure
also shows that, as expected, the transition to the asymmetric configuration
follows reducing $\kappa $.

\begin{figure}[th]
\centering \epsfig{file=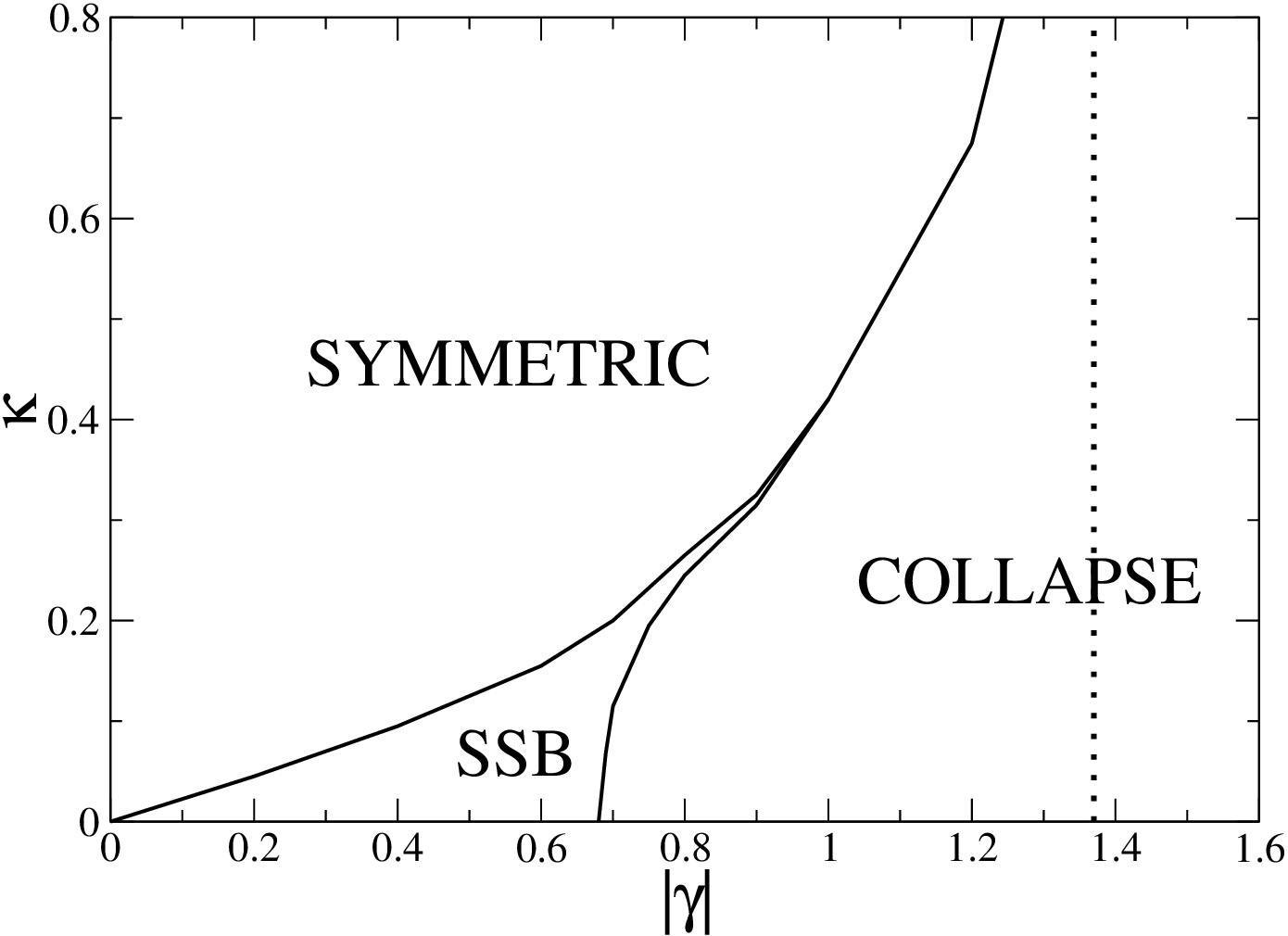,width=0.8\linewidth,clip=}
\caption{The phase diagram of the linearly-coupled system of Eqs. (\protect
\ref{npse-cyl}) for the vortex modes with $S=1$. In regions
\textquotedblleft symmetric" and \textquotedblleft SSB", the system
supports, respectively, stable symmetric [$u_{1}(r)=u_{2}(r)$] and
asymmetric [$u_{1}(r)\neq u_{2}(r)$] stationary solutions. The collapse
takes place in the eponymous region. The system always collapses on the
right side of the vertical dashed line. Units are the same as in Fig.
\protect\ref{fig1}.}
\label{fig5}
\end{figure}

For the modes with $S=1$, the phase diagram of the linearly-coupled system
of equations (\ref{npse-cyl}) in the parameter plane of $(\gamma ,\kappa )$
is displayed in Fig. \ref{fig5}. Comparing Figs. \ref{fig3} and \ref{fig5},
we conclude that the \textquotedblleft collapse region" is slightly reduced
at the nonzero vorticity. In this case, the system always collapses at $%
|\gamma |>1.36$.

\begin{figure}[th]
\centering
\epsfig{file=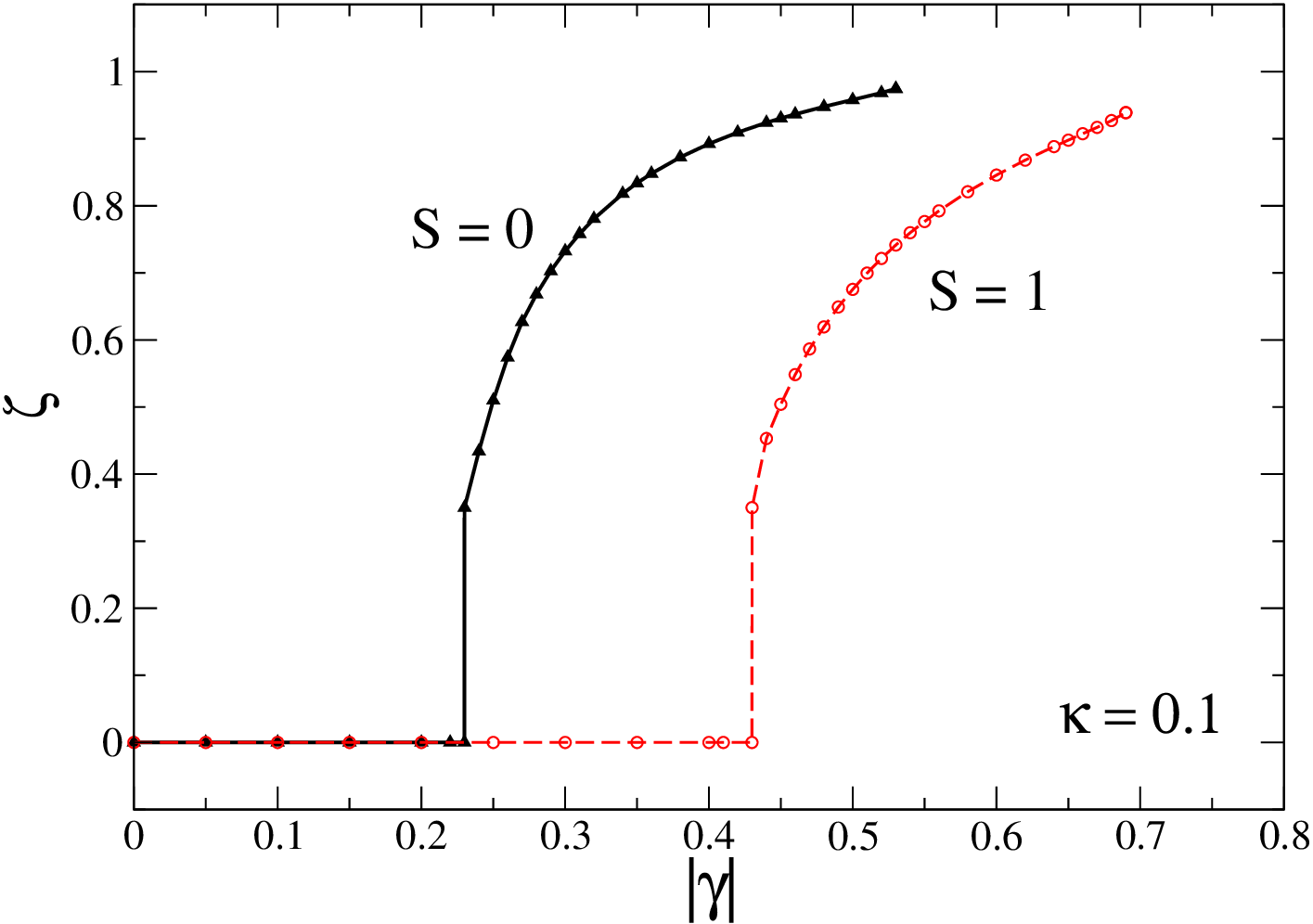,width=0.8\linewidth,clip=}
\caption{(Color online) The imbalance parameter, defined as per Eq. (\protect
\ref{zeta}), as a function of interaction strength $\protect\gamma $, for $%
S=0$ (filled triangles connected by the solid line), and $S=1$ (open circles
connected by the dashed line), for $\protect\kappa =0.1$. The curves
terminate at the collapse points. Units are as in Fig. \protect\ref{fig1}.}
\label{fig6}
\end{figure}

The SSB can be characterized by the imbalance (asymmetry) parameter,
\begin{equation}
\zeta (t)=\frac{N_{1}(t)-N_{2}(t)}{N_{1}(t)+N_{2}(t)}=\frac{N_{1}(t)-N_{2}(t)%
}{N}\;.  \label{zeta}
\end{equation}%
The competition between the symmetry breaking and collapse is further
illustrated in Fig. \ref{fig6} by plots of $\zeta $ versus $\gamma $ for a
relatively weak linear coupling, $\kappa =0.1$. In this figure, $\zeta $ is
the asymptotic value produced by the imaginary-time evolution in the
framework of Eqs. (\ref{npse-cyl}) with initial value $\zeta (0)=0.01$. The
curves in Fig. \ref{fig6} feature a leap (represented by vertical segments)
from the symmetric configuration with $\zeta =0$ to the asymmetric one with $%
\zeta \neq 0$. Actually, the transition to asymmetric states in the present
model always happens by a leap, i.e., the symmetry-breaking bifurcation is
always \textit{subcritical}, similar to the situation in the coupled
equations with the self-attractive cubic nonlinearity \cite{Warsaw}. The
figure shows that, at fixed $\kappa $, both the SSB and collapse happen at
higher values of $|\gamma |$ in the case of $S=1$,  with respect to $S=0$.

\subsection{Real-time dynamics}

\begin{figure}[th]
\centering
\epsfig{file=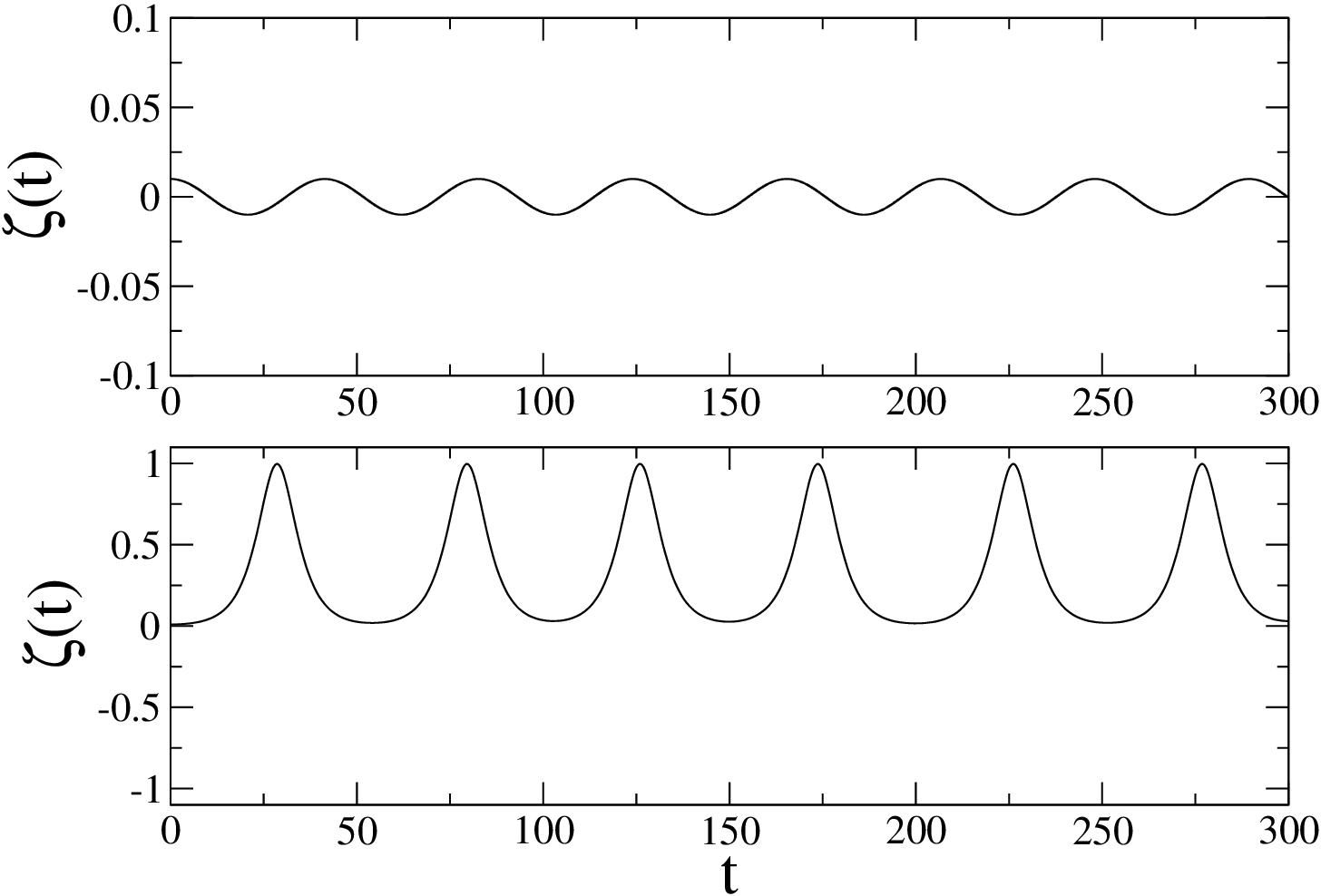,width=0.8\linewidth,clip=}
\caption{The real-time dynamics of the imbalance defined as per Eq. (\protect
\ref{zeta}), for $\protect\kappa =0.1$ and $S=0$. The upper and lower panels
correspond, respectively, to values of the interaction strength $\protect%
\gamma =-0.1$ and $\protect\gamma =-0.4$. The initial imbalance is $\protect%
\zeta (0)=0.01$. Units are as in Fig. \protect\ref{fig1}.}
\label{fig7}
\end{figure}

In the above subsections we have reported results of the imaginary-time
simulations, which produce the stationary solutions. The next step is to
test the stability of the modes by solving Eqs. (\ref{npse-cyl}) in real
time. In Fig. \ref{fig7} we display the real-time dynamics of the imbalance
parameter, $\zeta (t)$, of the system with $\kappa =0.1$ for $S=0$. The
initial value is $\zeta (0)=0.01$. In the upper panel, we chose $\gamma =-0.1
$, which corresponds to a stationary symmetric configuration, while in the
lower panel we set $\gamma =-0.4,$ which pertains to the asymmetric mode.

Figure \ref{fig7} shows that the dynamics are completely different in the
two cases. With the initial imbalance $\zeta (0)=0.01$ in both cases, $\zeta
(t)$ remains small in the course of the oscillations around the stationary
symmetric configuration, changing its sign periodically. Actually, $\zeta (t)
$ oscillates harmonically around the $\zeta =0$. In the case of the
stationary asymmetric configuration, the imbalance $\zeta (t)$ periodically
assumes very large values, but it does not change the sign; actually, $\zeta
(t)$ oscillates around a mean value, $\bar{\zeta}\simeq 0.5$. Note that the
value of $\zeta $ obtained asymptotically with these parameters ($\kappa =0.1
$ and $\gamma =-0.4$) in the imaginary-time simulations is $\zeta =0.89$,
see Fig. \ref{fig6}.

\begin{figure}[th]
\centering
\epsfig{file=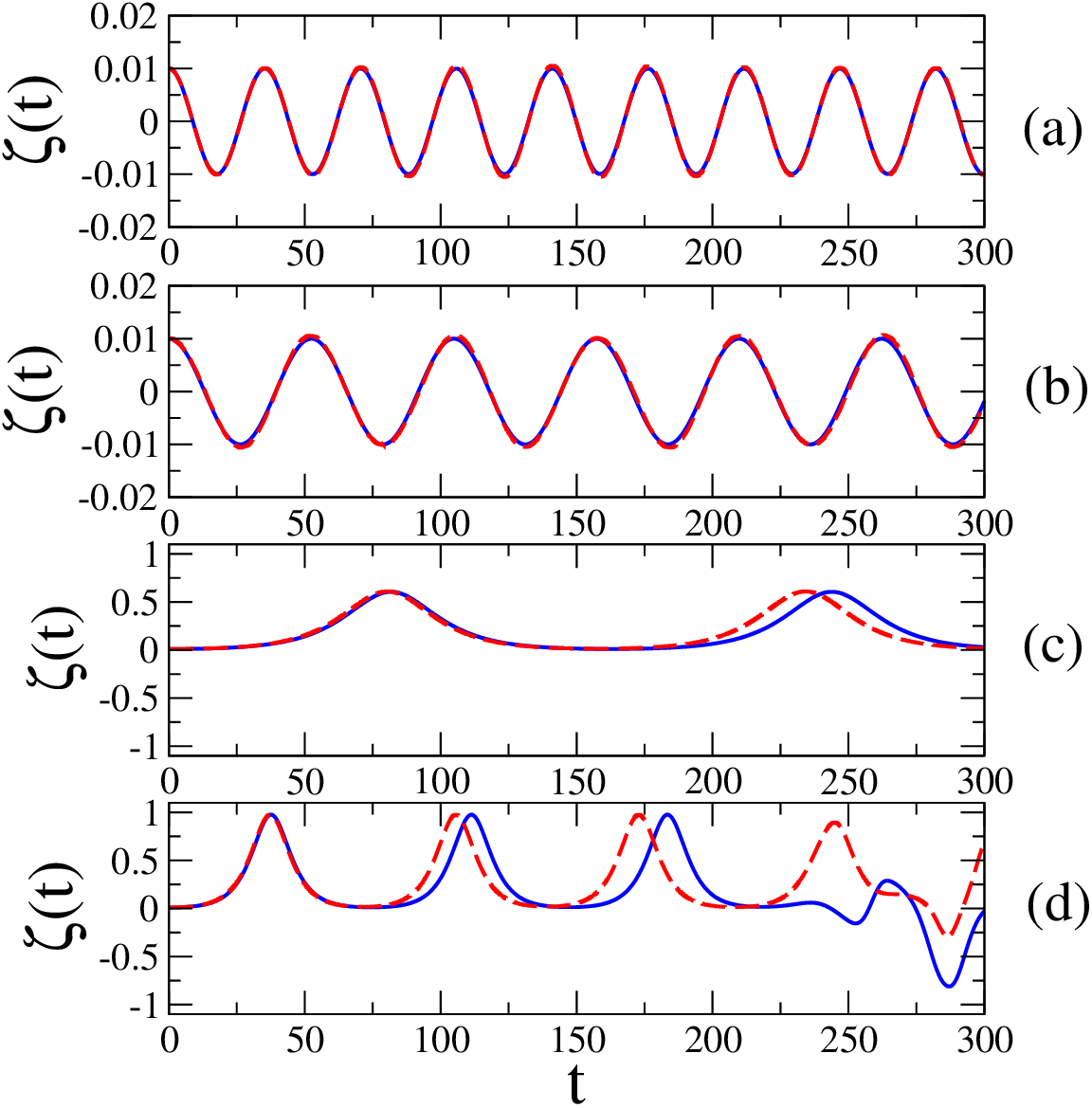,width=0.8\linewidth,clip=}
\caption{(Color online) The real-time dynamics of the imbalance, defined by
Eq. (\protect\ref{zeta}), for the vortex modes with $S=1$, at $\protect%
\kappa =0.1.$ Four values of the interaction strength $\protect\gamma $ are
considered: (a) $\protect\gamma =-0.1$, (b) $\protect\gamma =-0.3$, (c) $%
\protect\gamma =-0.5$, (d) $\protect\gamma =-0.7$. The initial imbalance is $%
\protect\zeta (0)=0.01$, in all the cases. The solid lines are obtained by
solving Eqs. (\protect\ref{npse}) for the axially-symmetric initial
conditions (\protect\ref{initial}) with $\protect\delta =1$, while the
dashed lines are generated by the initial conditions with $\protect\delta %
=1.1$, which break the azimuthal symmetry. Units are as in Fig. \protect\ref%
{fig1}.}
\label{fig8}
\end{figure}

In Fig. \ref{fig8} we plot the evolution of $\zeta (t)$ for $S=1$. It is
important to stress that when the stability of the vortex solutions is
tested against perturbations in real time, it is necessary to study the
stability of the vortex against azimuthal perturbations, which may lead to
splitting of the vortex that might seem stable in axially symmetric
simulations \cite{vor-stab}. For this reason, we employed full equations (%
\ref{npse}) to study the real-time dynamics of the vortices, considering
both axially-symmetric initial conditions and those breaking the azimuthal
symmetry. The general initial conditions used for the simulations of Eq. (%
\ref{npse}) are
\begin{eqnarray}
\Phi _{1}(x,y,t &=&0)=(x+iy)e^{-(x^{2}+\delta y^{2})/2}\;,  \notag \\
\Phi _{2}(x,y,t &=&0)=(x+iy)e^{-(x^{2}+\delta y^{2})/2}\;.  \label{initial}
\end{eqnarray}%
We set $\delta =1$ for the symmetric configurations, and $\delta =1.1$ for
ones with the broken azimuthal symmetry. Notice that $\Phi _{1}(x,y,t=0)$ is
normalized to $N_{1}(0)=1.01$ and $\Phi _{2}(x,y,t=0)$ to $N_{2}(0)=0.99$.
In Fig. \ref{fig8} the two upper panels [(a) with $\gamma =-0.1$ and (b)
with $\gamma =0.3$] correspond to stationary symmetric configurations (the
Josephson regime), with $\kappa =0.1$ (see Fig. \ref{fig6}). The results
displayed in these two panels of Fig. \ref{fig8} show (the solid lines
versus the dashed ones) that the additional azimuthal perturbation has no
appreciable effects in the dynamics, apart from a slight dephasing. The
third panel of Fig. \ref{fig8} [(c), with $\gamma =-0.5$] corresponds to a
stationary asymmetric vortex (in the self-trapping regime). Here we find
large-amplitude oscillations without a change in the sign of the population
imbalance, $\zeta (t)$. The solution with the unbroken azimuthal symmetry
(the dashed line) has a period of oscillations very close to that observed
in the solution with the azimuthal perturbation (the solid line). In any
case, we conclude that the asymmetric vortex with $\kappa =0.1$, $\gamma
=-0.5$ and $S=1$ is \emph{dynamically stable}. Finally, in the bottom panel
of Fig. \ref{fig8} we plot $\zeta (t)$ for parameters $S=1$, $\kappa =0.1$
and $\gamma =-0.7$, which are at the border of the collapse region (see Fig. %
\ref{fig6}). The panel shows that the solution is unstable (and eventually
suffers the collapse). Here, the main difference between solid and dashed
curves is the time after which $\zeta (t)$ displays the instability, which
may be identified as the instant at which $\zeta (t)$ changes its sign for
the first time. We observe that the instability with respect to the
azimuthal perturbations can produce an additional border inside both the
symmetric and asymmetric domains in Fig. \ref{fig5}. We did not aim to
produce this border in an exact form, as it is a computationally expensive
objective.

\section{Conclusions and open problems}

We have studied the dynamics of the self-attractive BEC in tunnel-coupled
disk-shaped traps, by means of systematic simulations of the coupled
nonpolynomial Schr\"{o}dinger equations derived from the 3D Gross-Pitaevskii
equation. In this way, we have investigated the phase diagram of the system
as a function of the interaction strength ($\gamma $) and tunneling
coupling. We have found that borders of different domains in the phase
diagram depend on vorticity $S$ of the localized modes: both the SSB
(spontaneous symmetry breaking) and collapse happen at larger values of $%
\gamma $ in the case of $S=1$ case with respect to the ground state ($S=0$).
We have also studied the dynamics of the two disk-shaped condensates around
the stationary configurations. Small-amplitude harmonic oscillations,
showing a periodic transfer of atoms between the condensates, take place
around the stable symmetric configurations. Instead, large-amplitude
oscillations without the change of the sign of the imbalance between the two
condensates occur around the perturbed asymmetric configurations.

There are many interesting open problems about Bose-Einstein condensates 
coupled by tunneling we want to face in the next future. In particular, 
we plan to investigate quasi one-dimensional and quasi 
two-dimensional Bose-Einstein condensates 
in nonlinear lattices (i.e. with space-dependent interaction strength)  
\cite{boris-nl} by using the nonpolynomial Schr\"{o}dinger equations. 
Moreover, we want to analyze the signatures of classical and quantum 
chaos \cite{sala-chaos} in these double-well configurations. Finally, 
we aim to calculate analytically the coupling tunneling energy 
of bosons by means of the WKB semiclassical quantization \cite{sala-wkb} 
and comparing it with the numerical results of the Gross-Pitaevskii 
equation. 

LS thanks Luciano Reatto for 9 years of fruitful scientific  
collaboration at the Physics Department of the University of Milano.

\end{document}